\documentclass[
    ,final            
  ]
  {aipproc}

\layoutstyle{6x9}

\newcommand\simlt{\lower.5ex\hbox{$\; \buildrel < \over \sim \;$}}
\newcommand\simgt{\lower.5ex\hbox{$\; \buildrel > \over \sim \;$}}

\begin{document}

\title{Phenomenology of Gamma-Ray Jets}


\author{Amir Levinson}{
  address={School of Physics and Astronomy, Tel Aviv University, Tel Aviv 69978, Israel}
}

\begin{abstract}
We discuss some phenomenological aspects of $\gamma$-ray emitting jets. 
In particular, we present calculations of the $\gamma$-sphere and $\pi$-sphere for various target photon fields and employ them to
demonstrate how $\gamma$-ray observations at very high energies can be used 
to constraint the Doppler factor of the emitting plasma and the production of VHE neutrinos.  We also
consider some implications of the rapid TeV variability observed in M87 and the TeV blazars, and
propose a model for the very rapid TeV flares observed with HESS and MAGIC in some blazars,
that accommodates the relatively small Doppler factors inferred from radio observations.  
Finally, we briefly discuss the prospects for detecting VHE neutrinos from relativistic jets.
\end{abstract}

\maketitle


\section{Introduction} 
Ejection of collimated relativistic outflows appears to be a common phenomena in astrophysics.
The radio-to-$\gamma$-ray continuum emission observed in blazars, microquasars, and $\gamma$-ray bursts (GRBs) is believed to
be produced in such outflows on various scales.  The common view is that these flows are powered by a magnetized accretion disk and a spinning black hole, and collimated by magnetic fields and/or the medium surrounding the jet.  However, there is as yet no universal agreement about the mechanisms responsible for the formation and 
acceleration the jet and the dissipation of its bulk energy.  Even the composition of the jet is unknown in most sources.  

Active states during which rapid, large amplitude variations of the high-energy emission are observed appear to be quite common in most classes of compact relativistic systems.  This activity is presumably associated with violent ejection episodes, as directly inferred in a few cases.  Some of the objects mentioned above may also provide sites for acceleration of the 
UHE cosmic rays detected by various experiments, if the latter are indeed produced in a bottom-up scenario, posing a great challenge to the theory of particle acceleration.
Emission of VHE neutrinos should accompany the production of those UHECRs, and optimistic models predict fluxes in excess of detection limit of upcoming cubic-km scale neutrino telescopes.  As discussed below,
observations of VHE $\gamma$-rays can provide stringent constraints on the photopion opacity, which can be translated into upper limits on the neutrino flux.

Of particular interest is the class of TeV sources.  There are at present over a dozen TeV blazars (for updated list see, e.g., Ref~\cite{wag07}), all of which are exclusively associated with the class of high peak BL Lac objects, another (nonblazar) TeV AGN, M87 \cite{ah06}, and several X-ray binaries (microquasars or gamma-ray binaries).  
The observed bolometric luminosity of TeV blazars during quiescent states is typically
of the order of a few times $10^{44}$ ergs s$^{-1}$, with about 10 percents emitted as VHE $\gamma$-rays. 
The luminosity in the VHE band may be larger by a factor of 10 to 100 during flaring states.
 The intrinsic spectra (corrected for absorption on the extragalactic background light) appear to be hard, 
with a peak photon energy in excess of 10 TeV in the most extreme cases.  The constraints on the dynamics of the system 
are most stringent in this class of sources, and are discussed in some greater detail below.

\section{Structure and Dynamics of $\gamma$-ray jets}
\subsection{The $\gamma$-sphere and the $\pi$-sphere}
The pair production opacity is typically large within the inner jet region in essentially all classes of compact high-energy sources.  
This implies that
$\gamma$-rays produced at small radii will not be able to escape the system before being converted to e$^\pm$ pairs.
Both the synchrotron photons produced inside the jet and ambient radiation intercepted by the 
jet contribute an opacity to pair production.  Results of detailed calculations of the pair production opacity are 
exhibited in figure 1, where the {\em $\gamma$-spheric radius}, defined as the radius 
$r_{\gamma}(\epsilon_{\gamma})$ beyond which the pair production optical depth to infinity
is unity, viz., $\tau_{\gamma\gamma}(r_\gamma,\epsilon_\gamma)=1$, is 
plotted against $\gamma$-ray energy $\epsilon_{\gamma}$, for two target radiation fields: 
synchrotron radiation (dashed lines) and external radiation (solid lines).  The spectra of the target
radiation fields employed in those calculations are given in Ref~\cite{lev06}.  As seen
the $\gamma$-spheric radius increases, quite generally, with increasing $\gamma$-ray energy, and for
luminous sources can be much larger than the dissipation radius (indicated in the figure).

In the powerful blazars, like 3C279, and in microquasars the intensity of both external and synchrotron radiation is quite large, 
corresponding to the uppermost curves in fig. 1.  This then implies that
the $\gamma$-spheres at energies corresponding to the GLAST band should encompass a rather large range of radii.  
In these sources the $\gamma$-spheres can be mapped in principle by measuring
temporal variations of the $\gamma$-ray flux in different energy bands during a flare.
If the $\gamma$-ray emission is produced over many octaves of jet radius, where 
intense pair cascades at the observed energies are important \cite{bln95}, then 
it is expected that a flare will propagate from low to high energies, or that the 
variations at higher $\gamma$-ray energy will be slower than at lower energies.
With the limited sensitivity and energy band of the EGRET instrument it was practically 
impossible to resolve such effects.  It is hoped that with the upcoming GLAST instrument this will be feasible.
In contrast, in TeV blazars, which are much fainter, the target photon luminosity is small and the location of the 
$\gamma$-spheres is not constrained, except, perhaps, at the highest energies observed (a few TeV).  This difference 
should be reflected in the variability pattern of the VHE emission.  A particular model for the rapid TeV flares is 
discussed below.
 \begin{figure}[h]
\includegraphics[height=.5\textheight]{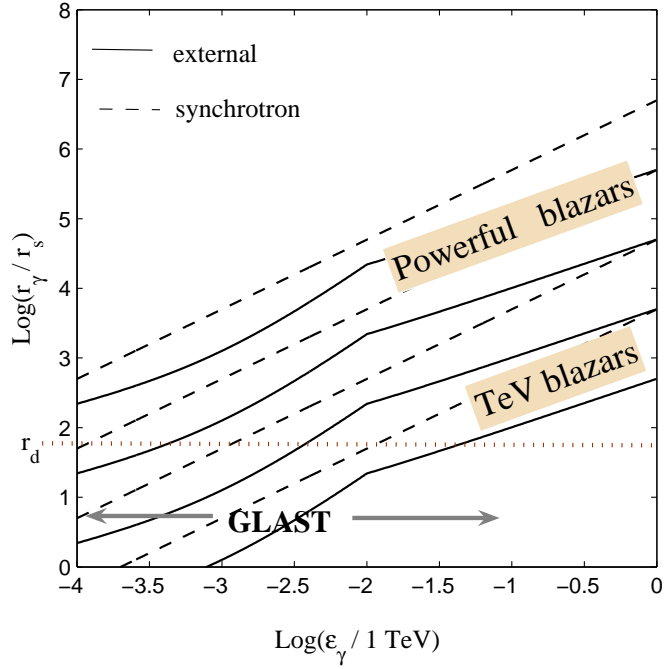}
\caption{Dimensionless $\gamma$-spheric radius versus $\gamma$-ray energy, computed in Ref~\cite{lev06}.  
The different curves correspond to a different normalization of the target radiation field intensity.
The dissipation radius $r_d$ is indicated.}
\end{figure}

The same target radiation field contributes also an opacity to photopion production.
Because both the protons and the $\gamma$ rays interact 
locally with the same target radiation field the ratio of photomeson and pair production opacities 
depend solely on the ratio of cross sections, $\sigma_{p\gamma}/\sigma_{\gamma\gamma}\simeq 4\times 10^{-3}$,
and the spectrum of the target radiation field.  For a target photon spectrum $n_s(\epsilon_s)\propto \epsilon_s^{-\alpha}$
we have \cite{lev06}
\begin{equation}
{\tau_{p\gamma}(\epsilon_p,r)\over \tau_{\gamma\gamma}(\epsilon_{\gamma},r)}
\simeq4\times10^{-3}
\left(\frac{\epsilon_p}{3\times10^5\epsilon_{\gamma}}\right)^{\alpha}.
\label{tau_pg}
\end{equation}
Detailed calculations of opacity ratios that employed more realistic target photon spectra
are presented in Ref~\cite{lev06}.   It is found that at $\gamma$-ray energies above 
a few TeV the opacity ratio is smaller than unity even at the maximum proton energy (determined from the confinement limit).  
For the class of TeV blazars this implies neutrino yields well below detection limit.
With GLAST it should be possible to constrain other sources and to use such constraints to identify the best candidates
for the upcoming km$^3$ detectors.

\subsection{Implications of variability}
The observed variability limits the linear size of the emission region (as measured in the Lab frame) to 
$d\simlt10^{14}(1-\beta\cos\theta_n)^{-1} t_{\rm var,h}/(1+z)$ cm, where $\beta$ is the bulk speed of the emitting 
fluid, $\theta_n$ is the viewing angle, $t_{\rm var,h}$ is the observed variability 
time in hours, and $z$ is the redshift of the source.  The rapid variability observed in GRBs and blazars implies
typically $d/r_g\simlt\Gamma^2$, where $\Gamma$ is the bulk Lorentz factor of the emitting fluid.  The rapid VHE flare
recorded recently in the TeV blazar PKS 2155-304 requires $\Gamma\sim20$ in order that $d\sim r_g$, regardless of
any other considerations.

The location of the emitting plasma is yet another issue.  If the emission originates from
radii $r_{\rm em}\sim d$, as often assumed, then compactness arguments yield a lower limit on the Doppler  
factor of the emission zone, as discussed further below.   If, on the other hand, the emission is produced
at radii $r_{\rm em}>>d$, as proposed recently for M87 and TeV blazars, then the fraction of jet energy 
that can be dissipated and converted to radiation in a conical jet of opening angle $\theta$ is at most
$\eta\simlt (d/\theta r_{em})^2$.  As a consequence, 
either the opening angle of the jet must be very small, $\theta \sim d/r<<1$, or the radiative efficiency $L_{\rm rad}/L_j$ 
must be very small, implying unreasonably large jet power in the most extreme cases.
In situations where the jet is underpressured relative to the confining medium the jet is expected to converge to 
the axis.   Reflection of collimation shocks at the nozzle may then give rise to appreciable dissipation 
in a very small region.   The pattern speed of the emission region can differ significantly from the speed of the fluid,
and stationary features, as occasionally observed in radio jets of blazars \cite{Jorstad01} can be naturally produced.
Such a model has been proposed to explain the rapid variability of the resolved X-ray emission and the unresolved TeV emission
from the HST1 knot in M87 \cite{cheu07}.  Alternative explanations have been offered for the rapidly varying TeV emission (e.g., Ref~\cite{m87}, and references therein).
It should be noted though that in M87 the X-ray and TeV luminosities, $L_{\rm TeV}\sim L_{\rm x}\simlt10^{41}$ erg s$^{-1}$ \cite{ah06,cheu07}, are much smaller than the TeV luminosity, $L_{\rm TeV}\sim 10^{44-45}$ erg s$^{-1}$, observed typically in the TeV blazars.  Estimates of the jet power in M87 
yield $L_j\simgt10^{44}$  erg s$^{-1}$ \cite{bick96}, implying a very small conversion fraction, $L_{\rm TeV}/L_j\simlt10^{-3}$.  Even with such a small conversion efficiency an opening angle $\theta<10^{-2}$ rad is required 
if the TeV emission were to originate from the HST1 knot, unless reconfinement can give rise to sufficient convergence of the jet at the location of HST1, as proposed in Ref~\cite{cheu07}.  This idea is compelling since even modest radiative cooling of the shocked jet layer in a proton dominated jet will lead to such a convergence, at least in the non-relativistic case \cite{eich82}.  
The effect of cooling on the collimation of relativistic jets needs to be explored.  The stationary radio features observed 
in blazars seem to indicate that recollimation shocks may be an important dissipation channel in blazsrs, and this may apply also to other sources, e.g., GRBs \cite{brom07}.  Whether the extreme TeV flares observed in VHE blazars can be accounted for by recollimation shocks at radii $r_{\rm em}>> d$ remains to be investigated.  However, this would not resolve the 'Doppler factor crises' if the IR emission would turn out to vary on timescales comparable to
the duration of the TeV flare.

\subsection{Constraints on Doppler factors}
Constraints on the Doppler factor of the fluid emitting VHE $\gamma$-rays can be derived by measuring the low-energy flux 
(radio-to-IR) simultaneously with the variable $\gamma$-ray emission.  
The requirement that the pair production opacity 
should not exceed unity, viz., $\tau^{\rm syn}_{\gamma\gamma}(r_{\rm em},\epsilon_\gamma/\delta)<1$,
constrains the density of target photons: $n_s(r_{\rm em})\simlt (\sigma_{\gamma\gamma} d)^{-1}$.
If the emission is assumed to originate from the innermost jet radii, in which case $r_{\rm em}\sim d$, then
we must have $r_{\gamma}(\epsilon_\gamma)< r_{\rm em}$.
The latter condition on $r_{\gamma}(\epsilon_\gamma)$ can be solved for the Doppler factor to yield \cite{lev06}
\begin{equation}
\delta^{5}>2\times10^{11}
t_{\rm var,h}^{-3/2}(\Gamma\theta)^{-2}z^2(\epsilon_\gamma/{\rm 1 TeV})^{1/2}S_{{\rm Jy}},
\label{Var-const}
\end{equation}
where $S_{\rm Jy}$ is the measured synchrotron flux density in Janskys and $z$ is the redshift of the source.
In deriving eq. (\ref{Var-const}) a luminosity distance $d_L=10^{28} z$ has been adopted.
In cases where  $r_{\rm em}>> d$ the compactness of the TeV emission zone may be constrained by the variability 
of the IR flux observed simultaneously with the TeV flare.   If the IR emission varies over time scales comparable to the
duration of the VHE flare then eq. (\ref{Var-const}) still applies.  Lower values of $\delta$ are allowed 
if the variability time of the IR emission is much longer than the variability time of the VHE emission.

Adopting $\Gamma\theta\simeq1$, we estimate $\delta>35$ for the rapid flare observed in Mrk 421 and $\delta>190$ 
for the few minuets variability reported for PKS 2155-304, with $r_{\rm em}<10^{17}$ cm for the minimum condition in both sources.

\subsection{A model for rapid flares in TeV blazars}
The large values of the Doppler factor implied by opacity constraints and rapid variability in TeV blazars 
are consistent with those obtained from fits of the SED to a homogeneous SSC model, but are in clear 
disagreement with the much lower values inferred from unification models \cite{Urry91,Hardcastle03} and 
superluminal motions on parsec scales \cite{mar99,Jorstad01,Giroletti04}.  
Various explanations, including a structure consisting of interacting spine and sheath \cite{Ghisellini05}, opening 
angle effects \cite{Gopal-Krishna04} and jet deceleration \cite{Geo03,Piner05} have been proposed in order to resolve 
this discrepancy. 

It has been proposed recently \cite{lev07} that the rapid TeV flares observed in sources like Mrk 421, Mrk 501 and PKS 2155-304 
are produced by radiative deceleration of fluid shells expelled during violent ejection episodes.
These shells are envisaged to accelerate to a Lorentz factor $\Gamma_0>>1$ at some radius $r_d\sim10^2-10^3 r_g$, at which dissipation of their bulk energy occurs.
The dissipation may be accomplished through formation of internal shocks in a hydrodynamic jet or dissipation of magnetic energy in a Poynting flux dominated jet \cite{rom92,lev98}, and it is assumed that a fraction $\xi_e$ of the total
proper jet energy density, $u_j^\prime$, is tapped for acceleration of electrons to a maximum energy $\gamma_{max}m_ec^2$.  
The dynamics of the front is then governed by the equation \cite{lev07}
\begin{equation}
\frac{d}{dr}(u^\prime_j\Gamma^2 \beta)= -\frac{4\sigma_T}{3m_ec^2}\chi\xi_e \Gamma^3\gamma_{\rm max}u_su^\prime_{j},
\label{eq-mot}
\end{equation}
where $u_s$ is the energy density of the target radiation field, as measured in the Lab frame, and $\chi=<\gamma^2>/(<\gamma>\gamma_{\rm max})$ depends on the energy distribution of nonthermal electrons.
For a power law distribution, $dn_e/d\gamma\propto \gamma^{-q}$ with $q\le2$, we have $1>\chi>\>0.1$.
Under the assumptions that $u_s(r)\propto r^{-2}$ and that the proper density and average energy of 
the nonthermal electrons are independent of radius the solution of eq. (\ref{eq-mot}) (in the limit $\beta=1$) reads:
\begin{equation}
\Gamma_\infty=\Gamma_0 \frac{l}{l+r_d},
\end{equation}
where $\Gamma_\infty$ is the asymptotic Lorentz factor downstream.  The stopping length can be expressed in terms 
of the optical depth for $\gamma\gamma$ absorption of a $\gamma$-ray of energy $m_ec^2\epsilon_{\gamma}$ 
by a power law target photon field of the form  $I_s(\epsilon_s)\propto \epsilon_s^{-\alpha}$; $\epsilon_{s,min}<\epsilon_s<\epsilon_{s,max}$, as
\begin{equation}
\frac{l}{r_d}=\frac{1}{\chi\xi_e\tau_{\gamma\gamma}}\left(\frac{\sigma_{\gamma\gamma}}{\sigma_T}\right)
\left(\frac{\epsilon_\gamma}{\Gamma_0\gamma_{\rm max}}\right)g(\epsilon_\gamma),
\label{stopp2}
\end{equation} 
with $g(\epsilon_{\gamma})=(\epsilon_\gamma\epsilon_{s,min})^{\alpha-1}$ if $\alpha>1$ and  
$g(\epsilon_{\gamma})=(\epsilon_\gamma\epsilon_{s,max})^{\alpha-1}$ if $\alpha<1$, and 
$g(\epsilon_\gamma)\le1$ in both cases.  We conclude that for a reasonably flat distribution of nonthermal electrons, $q\le2$, extension of the distribution to a maximum energy $\gamma_{\rm max}$ at which the pair production optical depth, $\tau(\Gamma_0\gamma_{\rm max})$, is a few is already sufficient to cause appreciable deceleration of the front.

From the above it can be shown \cite{lev07} that for the TeV blazars a background luminosity of $L_s\sim 10^{41}-10^{42}$ erg s$^{-1}$, roughly the luminosity of LLAGN, would lead to a substantial deceleration of the front 
and still be transparent enough to allow the TeV $\gamma$-rays produced by Compton scattering of the background photons
to escape the system.  The ambient radiation field is most likely associated with the nuclear continuum source.  The bulk Lorentz factor of the jet during states of low activity may be appreciably smaller than that of fronts expelled during violent ejection episodes.


\section{What is the prospect for detection of VHE neutrinos from relativistic jets?}
A new generation of experiments just started operating or will become operative soon, design to detect
VHE neutrinos (IceCube, ANTARES, NESTOR, NEMO),
and UHE cosmic rays (HiRes and the hybrid Auger detectors), are probing and will probe regions opaque to 
electromagnetic radiation, which are presently unaccessible, and will determine the composition of jets.
Besides providing an important probe of the 
innermost regions of compact astrophysical systems, these experiments can also be exploited to 
test new physics.  

As mentioned above, the detection of UHECRs motivated considerations of heavy jets that effectively accelerate protons to 
energies approaching the confinement limit.   Effective neutrino production requires large photopion opacity which,
as discussed above, can be constrained by VHE $\gamma$-ray observations.   Present observations rules out
TeV blazars as potential candidates for the upcoming km$^3$ neutrino telescopes (see Ref~\cite{lev06} for further
discussion).  This leaves GRBs, microquasars, and perhaps powerful EGRET blazars as potential candidates.
Powerful blazars at a redshift of $z=1$ may produce up to one event per year in a km$^3$ detector \cite{Atoyan01}.  In the case of microquasars 
up to a few events can be detected during a strong outburst if the viewing angle is sufficiently small 
\cite{levw01,dest02,torr07}.   The estimated neutrino flux from GRBs implies that only nearby sources can be individually detected by the 
upcoming experiments.  However, the cumulative flux produced by the entire GRB population should be detectable
assuming that cosmological GEBs are the sources of the observed UHECRs \cite{wax95}.



\begin{theacknowledgments}
This work was supported by an ISF grant for the Israeli Center for High Energy Astrophysics. 
\end{theacknowledgments}

\end{document}